\documentclass[format=acmsmall,review=false,screen=true]{acmart}

\usepackage{booktabs} 						
\usepackage{graphicx}
\usepackage{subcaption}

\renewcommand\footnotetextcopyrightpermission[1]{} 
\setcopyright{none}

\begin{document}
\title[Understanding the Production and Consumption of Clickbaits]{Tabloids in the Era of Social Media? Understanding the Production and Consumption of Clickbaits in Twitter}  

\author{Abhijnan Chakraborty}
\affiliation{%
  \institution{Indian Institute of Technology Kharagpur, India}
}

\author{Rajdeep Sarkar}
\affiliation{%
  \institution{Indian Institute of Technology Kharagpur, India}
}

\author{Ayushi Mrigen}
\affiliation{%
  \institution{Indian Institute of Technology Kharagpur, India}
}

\author{Niloy Ganguly}
\affiliation{%
  \institution{Indian Institute of Technology Kharagpur, India}
}

\begin{abstract}
With the growing shift towards news consumption primarily through social media sites like Twitter, most of the traditional as well as new-age media houses are promoting their news stories by tweeting about them. The competition for user attention in such mediums has led many media houses to use catchy sensational form of tweets to attract more users -- a process known as {\it clickbaiting}. In this work, using an extensive dataset collected from Twitter, we analyze the social sharing patterns of clickbait and non-clickbait tweets to determine the organic reach of such tweets. We also attempt to study the sections of Twitter users who actively engage themselves in following clickbait and non-clickbait tweets. Comparing the advent of clickbaits with the rise of {\it tabloidization} of news, we bring out several important insights regarding the news consumers as well as the media organizations promoting news stories on Twitter. 
\end{abstract}


\maketitle

\section{Introduction}
\label{sec:intro}
Historically, news has been a very important part of societal evolution, playing an integral role in the same since the 17th century~\cite{stephens2007history}. In the years that followed, the medium has only increased the power it wields over the perceptions and beliefs of the general public. The press has been considered as one of the fundamental pillars of any functioning democracy for a considerable period of time, and the editors of news organizations have long viewed their roles as the custodians of society, deciding which news should be consumed by the common people, and which shouldn't~\cite{shoemaker2009journalists}.

The first disruption in the media landscape came in the late 20th century from the {\it tabloidization of news}, where detailed truthful reporting of important events gave way to stories that were more sensational in nature. 
Since its inception, the tabloid style has been criticized 
as inferior, appealing to base instincts and public demand for sensationalism~\cite{bird2009tabloid}.

In the 21st century, online news consumption has gained momentum, and the revenue model has shifted towards being mostly advertisement driven, where users do not have to pay anything to read the news articles and the money comes from the clicks on the advertisements that are present on the news websites. Such shift in news consumption has caused a significant change in both the consumption pattern as well as the means of offering news content to the readers. 

The attractiveness of a news article's content has become more relevant than the credibility of the organization writing that article, and news readers now have a negligible cost for switching from one media source to another. This, in turn, has resulted in a fierce competition between the news outlets to capture the readers' attention. A by-product of this competition is the advent of {\bf clickbaits}, where in order to tempt the readers to click and read articles in their websites, the media organizations use flashy headlines that pique the interest of the readers~\cite{chakraborty2016stop}. 

Clickbaits are typically defined as the headlines that are intended to lure readers, by providing a small glimpse of what to expect from the article~\cite{blom2015click}. According to the Oxford English Dictionary, clickbait is {\it ``(On the Internet) content whose main purpose is to attract attention and encourage visitors to click on a link to a particular web page"}
~\footnote{\url{oxforddictionaries.com/us/definition/american_english/clickbait}}. 
Examples of clickbaits include ``17 Reasons You Should Never Ever Use Makeup", ``These Dads Quite Frankly Just Don't Care What You Think", or ``10 reasons why Christopher Hayden was the worst `Gilmore Girls' character". 

On one hand, the success of such clickbait headlines in attracting visitors to the news websites has propelled new age digital media companies like BuzzFeed to have valuation 
three times that of century old The Washington Post
~\cite{buzzfeed_valuation}. 
However, on the other hand, the articles themselves often offer less news value, and therefore, concerns have been raised by traditional media organizations regarding the role of {\it journalistic gatekeeping} in the era of clickbaits
~\cite{clickbait_bad,clickbait_bbc}.

The rise of clickbaits in online news bears many similarities with the advent of tabloids. In general, there has been a lot of discussion about the positive and negative effects of tabloidization in the traditional media. Many researchers believe that tabloids have had a significant role to play in lowering the standards of news. According to Colin Sparks, {\it ``Public ignorance and apathy is growing as the serious, challenging and truthful is being pushed aside by the trivial, sensational, vulgar and manipulated"}~\cite{kevin2003media}.  However, on the other hand, other researchers have argued that {\it softening} of news by the tabloids helped raising political awareness among politically inattentive citizens~\cite{baum2006oprah,delli2001let}. 
There have been long drawn debates around the content of tabloids, and the effects of the same on the {\it public sphere}~\cite{ornebring2004tabloid}. 
In spite of all the hue and cry against tabloidization, it has been agreed that blindly disregarding the importance of the tabloid form of journalism is not advisable~\cite{ornebring2004tabloid}.

Contrary to the study of tabloids, research on clickbaits is still in a very nascent stage, where recently a few attempts have been made to automatically detect clickbait headlines in different media sites~\cite{chakraborty2016stop,anand2017we,downworthy,potthast2016clickbait}. 
However, to our knowledge, there has been no attempt to understand the consumers who follow such headlines, 
and further help in expanding the reach of the corresponding articles to a wider audience. 
In fact, all past works mostly highlight the negative aspects of clickbaits without considering their consumers~\cite{chakraborty2016stop,chen2015misleading}. 

Considering the consumers of news is especially important today because there has been another parallel shift in news consumption over the last few years, where social media has become the primary medium for news consumption~\cite{Facebook-TV-Viewership-Survey}. 
This has lead to two other disruptions for the media organizations. 

First, in social media, anyone can become a publisher of content with virtually zero upfront cost, which has lead to mushrooming of several {\it social media only} publisher startups~\cite{social-media-only-news}. 
Emergence of such organizations has further crowded the media landscape already flooded with national, international and local news outlets, 
resulting in immense competition for user attention in such mediums. 

Second, the media organizations also need to adapt to another news dissemination medium,  
specifically related to the posting rules of that particular medium. A prime example of this is Twitter, which severely limits the number of characters that a single tweet can contain, and thus imposing an upper bound on the number of words that the news outlets can use to attract the users. 

Such drastic changes in the medium of propagation of news, and the competition for user attention in such mediums have led many media houses to use catchy sensational posts to attract more users. We broaden the original definition of clickbaits to include such catchy social media posts, which not only encourage users to click on the embedded news article links, but also persuade users to share these posts with their peers, which in turn help in increasing the media house's follower base.  
With the growing adoption of clickbaiting techniques, and social news consumption leading to transformational changes in the media landscape, we believe that at this point, a detailed study of clickbaits and their consumers in social media is required to get the holistic picture.

In this paper, we take a step towards that direction by analyzing the users as well as the usage of clickbaits in social media sites.   
We collect extensive longitudinal data over eight months from the popular social media Twitter covering both clickbait and non-clickbait (or traditional) tweets, 
and then attempt to explore several important dimensions related to these two types of tweets and their users.

More specifically, in this paper, we attempt to investigate the following research questions: \\
\noindent \textbf{RQ1. How are clickbait tweets different from non-clickbait tweets?} \\
\noindent \textbf{RQ2. How do clickbait production and consumption differ from non-clickbaits?} \\
\noindent \textbf{RQ3. Who are the consumers of clickbait and non-clickbait tweets?} \\
\noindent \textbf{RQ4. How do the clickbait and non-clickbait consumers differ as a group?}

Our investigation reveals several interesting insights on the production of clickbaits. 
For example, clickbait tweets include more entities such as images, hashtags, and user mentions, 
which help in capturing the attention of the consumers. 
Additionally, we find that a higher percentage of clickbait tweets convey positive sentiments as compared to non-clickbait tweets. 
As a result, clickbait tweets tend to have a wider and deeper reach in its consumer base than non-clickbait tweets. 

We also make multiple interesting observations regarding the consumers of clickbaits. For example, clickbait tweets are consumed more by women than men, as well as by more younger people compared to the consumers of non-clickbaits. Additionally, they have higher mutual engagement among each other. On the other hand, non-clickbait consumers are more reputed in the community, and have relatively higher follower base than clickbait consumers. 

Comparing clickbaits with tabloids, we find commonalities (e.g., both make optimal use of graphic elements) as well as differences (e.g., clickbaits tend to convey more positive sentiments vs more negativity in tabloid reporting). Likewise, clickbait and tabloid readership have similarities (e.g., both cater to younger population) and differences (e.g., majority of tabloid readers are male, which is in contrast to the female dominance among clickbait consumers). 

In summary, we make two major contributions in this paper: (i) to our knowledge, ours is the first attempt to understand the consumers of clickbaits, and (ii) while doing so, we also make the first effort to contextualize the rise of clickbaits with the tabloidization of news. As we  mentioned earlier, all past works on clickbaits only highlight its negative aspects. We believe that our work can foster further research, raise debates in the community, and help bring in a more holistic view of the entire spectrum.
\vspace{-2mm}
\section{Background and Related Work}
\label{sec:related}

\subsection{Tabloidization and its impact on news media}
Clickbaits can be thought as the digital successor to the {\it tabloidization} of print journalism~\cite{skovsgaard2014tabloid}. 
Tabloids disrupted a long-held approach towards {\it journalistic gatekeeping} by focusing more on {\it soft news} than 
{\it hard news}, and on sensationalizing the content over the detailed truthful reporting of events. There have been concerns 
in the journalism community regarding the tabloidization of news and its potential threat to democracy~\cite{rowe2011obituary,skovsgaard2014tabloid}. However, on the other hand, several studies have noted that {\it softening} of news by the tabloids helped raising political awareness among politically inattentive citizens~\cite{baum2006oprah,delli2001let}. 

Likewise, with the sudden increase in the prevalence of clickbaits in the digital media landscape, similar concerns have been raised. There have been different discussions constantly reprimanding the low news value of clickbaits, and the change in the face of journalistic gatekeeping that clickbaits bring with them
~\cite{clickbait_bad}. 
In this paper, we try to tie the existing works on tabloids, and see whether some of the arguments made there still hold for clickbaits. We also attempt to argue for a holistic debate abound the usefulness of clickbaits, similar to the one present for tabloids.

\subsection{Readership of traditional and tabloid newspapers}
Traditionally, to decide the correct audience for advertising goods and services, advertisers as well as market research 
agencies conducted readership surveys for offline newspapers. In such surveys, 
it has been found that the traditional broadsheet newspapers mostly cater to the affluent and well educated
audience, with majority of readers in the upper middle class and middle class professional and managerial social grades (denoted as grade AB by  National Readership Survey\footnote{\url{nrs.co.uk/nrs-print/lifestyle-and-classification-data/social-grade}} in the UK)~\cite{johansson2007reading}. 
Whereas, the majority of the tabloid readers are found to be in the C1C2-E social grades, i.e., people in lower middle and working class, 
 involved in supervisory, clerical, or skilled and unskilled manual works~\cite{rooney2000thirty}. Thus, there seems to be a clear distinction between the consumers of tabloids and traditional news. Following this line of work, in this paper, we investigate whether the clickbait and non-clickbait consumers in Twitter also differ substantially.

\subsection{Tabloids and the public sphere}
Most of the criticisms around tabloidization are grounded in the notion of the {\it public sphere}, developed in the seminal work by Habermas~\cite{habermas1991structural}. According to Habermas, public sphere is a realm of our social life where public opinion can be formed via {\it rational-critical debate between private individuals on public matters}~\cite{habermas1991structural}, and news media is the primary enabler of  such communications. By such construction, Habermas puts media in the normative center of a well-functioning democracy, and hence the standard of the content propagated by the media becomes immensely important~\cite{eide1999public}. Critiques of tabloids have argued that  tabloids fail those standards to enable debates in the public sphere~\cite{johansson2007reading}.
In this work, we investigate a related question in the context of social media -- among clickbaits or non-clickbaits, which is a better enabler of  communication between different groups of people in the social media, which is the public sphere in this context.

\subsection{Social media and journalistic gatekeeping}
The wide-spread adoption of social media sites like Facebook and Twitter has led to a paradigm shift in how news stories are consumed by people world-wide. Today, a large and growing fraction of news readers are finding news stories on these social media platforms. Unlike the traditional media landscape, the newspaper editors no longer exert editorial control over the type of stories getting shared, and hence, there is little journalistic gatekeeping applied on the news discourse. 

A lot of prior works have discussed the effect of such drastic change in the news landscape. By studying the sharing patterns of $8,000$ news items across major news outlets, Diakopoulos {\it et al.}~\cite{diakopoulos2014newsworthiness} concluded that the newsreaders in social media act as {\it network gatekeepers}, and more often (re)share news items on socially deviant events. Chakraborty {\it et al.}~\cite{chakraborty2016dissemination} analyzed how such network gatekeeping role differs across different sharing mediums. In another work, they observed that the gatekeeping roles exercised by both traditional and network gatekeepers can lead to temporal coverage bias~\cite{chakraborty2015can}. Further, Matias {\it et al.}~\cite{matias2017followbias} found gender bias among the network gatekeepers, where men get much greater attention compared to women. The authors also designed approaches to mitigate this gender bias. By analyzing the sharing behaviors of network gatekeepers, Orellana-Rodriguez {\it et al.}~\cite{orellana2016spreading} proposed a set of guidelines to maximize engagement with the shared news items. Complementary to all these prior works, in our present work, we attempt to distinguish between the network gatekeepers who consume and share clickbait and traditional news stories.

\subsection{Curiosity gap and psychological appeal of clickbaits}
Media organizations have always struggled to bridge the gap between what the news producers tend to promote, and what the news readers are actually choosing to read. Some examples include journalists' penchant for public affairs while readers showing a reduced interest in the same. To overcome such problems, news producers often tend to use psychological methods to appeal the readers. The methods include creating {\it curiosity gap}~\cite{loewenstein1994psychology} in the headline of the news article itself. Few recent works have attempted to understand such psychological appeals of clickbaits.

Specifically, researchers have examined how clickbaits employ two forms of forward referencing -- {\it discourse deixis} and {\it cataphora} -- to lure the readers to click on the article links~\cite{blom2015click}. Typically, this includes using pronouns to make reference to forthcoming parts in the discourse (often in the article text). For instance, the clickbait headline ``This Twitter User Says She Was Suspended After Criticising Taylor Swift" is cataphoric because here `This' refers to the name of the Twitter user, which is revealed only in the article body. In such headlines, there remain empty slots in the readers' mind that cannot be filled without going through the article in question~\cite{blom2015click}. These psychological tactics that clickbaits are known to engage, have become the crux of the arguments in favour of labeling clickbaits as misleading content or false news~\cite{chen2015misleading}.

\subsection{Automatic detection of clickbaits}
There have been some recent attempts to detect and prevent clickbaits. Facebook attempted to remove clickbaits depending on the click-to-share ratio and the amount of time spent on different stories~\cite{facebook_clickbait}. 
The browser plugin `Downworthy'~\cite{downworthy} detects clickbait headlines using a fixed set of common clickbaity phrases, and then converts them into meaningless garbage. Potthast {\it et al.}~\cite{potthast2016clickbait} attempted to detect clickbaits in Twitter 
by using common words occurring in clickbait tweets. Biyani {\it et al.}~\cite{biyani20168} proposed approaches to detect clickbaits using article informality. Anand {\it et al.}~\cite{anand2017we} used deep learning based techniques to detect clickbaits.

In our prior work~\cite{chakraborty2016stop}, we compared clickbaits and traditional news headlines, and noticed that clickbait headlines use several language traits to attract users. For example, such headlines have more function words, more stopwords, more hyperbolic words, more internet slangs, and more frequent use of possessive case, as compared to the traditional headlines where the title contains specific proper nouns and the reporting is in third person~\cite{chakraborty2016stop}. Based on these observations, we developed a clickbait classifier where given a news article headline, the classifier would classify it as clickbait or non-clickbait. In our present study, we extend the classifier developed in~\cite{chakraborty2016stop} to separate clickbait and non-clickbait tweets. 

However, the research questions we investigate in this work are complementary to the earlier work. For example, in~\cite{chakraborty2016stop}, we identified linguistic characteristics that differentiate clickbait and traditional news headlines. Whereas, in this paper, we explore complementary questions specific to tweets such as whether clickbait tweets contain several entities which might lead to their increased visiblity, or whether the sentiment conveyed by the clickbait tweets differ from the non-clickbait tweets. Moreover, taking a very different direction compared to~\cite{chakraborty2016stop}, we study the production and consumption patterns of clickbaits in Twitter, and bring out interesting insights.
\section{Dataset Gathered}
\label{sec:dataset}
As mentioned earlier, in this work, we attempt to analyze the usage as well as the users of clickbaits in social media. 
Towards that end, we gathered extensive longitudinal data from Twitter, covering a period of 8 months from February, 2016 to 
September, 2016. Throughout this 8 month period, using the Twitter Streaming API\footnote{\url{dev.twitter.com/streaming/overview}}, 
we collected all tweets posted by the Twitter handles of several (i) traditional news media organizations, and (ii) digital media outlets known to often deploy clickbaits. 

As news media organizations, we considered the top three newspapers according to the Alexa ranking\footnote{\url{alexa.com/topsites/category/News/Newspapers}}: New York Times, Washington Post and India Times. Additionally, we considered one {\it online only} news media outlet -- Huffington Post.
Interestingly, these media organizations do not maintain a single Twitter account. Rather, alongside the primary account (i.e., @nytimes,
@washingtonpost, @indiatimes, and @HuffPost), they also maintain several secondary accounts to tweet about stories related to specific news sections. For example, New York Times maintains more than $20$ Twitter accounts (e.g., @nytpolitics, @nytnational, @nytimesworld, @nytopinion, @nytimesbusiness etc.). Similarly, Washington Post maintains more than $10$ Twitter accounts (e.g., @postpolitics, @PostWorldNews, @PostOpinions, @PostSports etc.).  In total, we collected tweets posted by the four primary accounts, and $38$ secondary Twitter accounts of the media organizations in this category. 

Regarding the media outlets promoting clickbaits, we considered the five outlets identified in our earlier work~\cite{chakraborty2016stop}: BuzzFeed, Upworthy, ViralNova, ScoopWhoop, and ViralStories. Additionally, we identified three more outlets:  MensXP, 9GAG, and CountryLiving. We collected the tweets posted by their corresponding Twitter handles: @BuzzFeed, @Upworthy, @ViralNova, @ScoopWhoop, @allviralstories, @MensXP, @9GAG, and @CountryLiving. 
Similar to the traditional media organizations, many of these outlets also maintain multiple Twitter handles. For example, BuzzFeed maintains  @BuzzFeedPol, @BuzzFeedNews, @BuzzFeedFashion etc. alongside @BuzzFeed. Similarly, Scoopwhoop maintains several secondary Twitter accounts like @ScoopwhoopONN, @scoopwhoopnews, @ScoopWhoopVideo etc. In total, we collected the tweets posted by $27$ Twitter handles, which include the primary accounts as well as the secondary accounts of the media outlets in this category.

\begin{table}[t]
\small
\begin{center}
\begin{tabular}{|p{0.2\textwidth}||p{0.75\textwidth}|}
\hline
{\bf Twitter Handle} & {\bf Example Clickbait Tweets} \\
\hline
@BuzzFeedMusic &	 This theory about the new Radiohead album is driving fans crazy https://t.co/W26Glw5pEQ https://t.co/N9q5kIgsax  \\
\hline
@HuffingtonPost &	 39 breastfeeding portraits that celebrate nursing mamas https://t.co/D2XKXBeVmY https://t.co/PG48j7Dbuw \\
\hline
@BuzzFeed &	 Can You Guess Which One Of These People Is Holding A Vibrator? https://t.co/AhWF45yhzA \\
\hline
@ScoopWhoop &	 One of these places is in \#Switzerland. Can you guess which one it is?:  https://t.co/7VRufhbaMp https://t.co/nuh5n9xSnR \\
\hline
@Upworthy &	 These are 7 things they don't tell you about living with PCOS. https://t.co/CbwJSjtw4K https://t.co/GHtmKu5lIw \\
\hline
\hline
{\bf Twitter Handle} & {\bf Example Non-clickbait Tweets} \\
\hline
@HuffPostPol & Texas Lt. Gov. blames Black Lives Matter, social media for Dallas shooting https://t.co/wZRA0tId5e https://t.co/7v5y3NzUAU \\
\hline
@NYTNational & Milwaukee joins a growing list of cities to experience police-related racial violence. https://t.co/ymPdXrVhNz https://t.co/CbZkfA2J4B \\
\hline
@nytimesworld & How Kurds backed by the Pentagon ended up fighting Syrian Arabs backed by the CIA in Syria https://t.co/on3XgfhLnl https://t.co/JnkxTg4kgQ \\
\hline
@BuzzFeedWorld &	 EgyptAir Flight Carrying 69 People Disappears From Radar https://t.co/XrRFYEIPCP \\
\hline
@nytpolitics & U.S. Strike in Yemen Kills Dozens in Qaeda Affiliate, Officials Say https://t.co/PXhABcCdLF  \\
\hline
\end{tabular}
\caption{{\bf Few examples of clickbait and non-clickbait tweets in our dataset.}}
\vspace{-5mm}
\label{tab:example}
\end{center}
\end{table}

For both traditional media organizations and the outlets promoting clickbaits, in addition to the original tweets posted by them, we also collected all retweets (of these original tweets) made by their followers on Twitter. In total, we collected around $288$K original tweets and $11.4$M retweets posted during February, 2016 to September, 2016. Then, we attempted to separate the data into two categories: (i) a  corpus of clickbait tweets, and (ii) another corpus of non-clickbait tweets. 

In this work, we used a slightly modified version of the clickbait classifier developed in~\cite{chakraborty2016stop}. In~\cite{chakraborty2016stop}, we identified several linguistic features which distinguish clickbait and traditional news headlines. These features include the length of the headline, the ratio between the number of stopwords to the number of content words, length of the longest separation between syntactically dependent words, the presence of cardinal numbers in the beginning of the headline, or the presence of unusual punctuation patterns and contracted word forms. Additionally, common clickbait phrases, internet slangs and determiners, word N-grams, POS N-grams and Syntactic N-grams were also used as features for the SVM classifier distinguishing clickbaits and non-clickbaits.

In order to apply the classifier to classify tweets, we removed the hashtags and urls from the tweets, and only considered the tweet text for classification. We also excluded a particular feature used in the original classifier: length of the news headline (here, number of words in a tweet). In Twitter, all tweets are restricted to $140$ characters, and therefore, we didn't find the length feature to be a major distinguisher  between clickbait and non-clickbait tweets. We retained all other features used in the original classifier, and applied it to separate clickbait and non-clickbait tweets.

To evaluate the performance of the classifier, we asked three volunteers to manually label $200$ randomly selected tweets from the dataset as either clickbait or non-clickbait. The inter-annotator agreement was `substantial' with Fleiss' $\kappa$ score~\cite{fleiss1971measuring} of $0.72$. Taking the majority vote as ground truth, $88$ tweets were identified as clickbait and rest of the tweets as non-clickbait. Then, we used the classifier, as described above, to classify the same tweets into clickbait and non-clickbait categories. 
By comparing the class predicted by the classifier, and the ground truth labels assigned by the human volunteers, we find the precision of the classification to be $0.872$, recall $0.843$, and F1-score to be $0.857$. 
The accuracy of the classifier was $87.6\%$ (against the random guessing accuracy of $50\%$).
 
As mentioned earlier, we applied this classifier on the tweets in our dataset (after removing hashtags and urls present in the tweets), and 
separated the data in two corpuses. The corpus of clickbait tweets consists of $126$K original tweets, 
and more than $5$M retweets posted by $1.08$M unique users.
On the other hand, the corpus of non-clickbait tweets contains $162$K original tweets 
and $6.4$M retweets posted by $1.29$M unique users. 
Table~\ref{tab:example} shows few example tweets from both clickbait and non-clickbait corpus. 
There are around $34$K users who retweeted both clickbait and non-clickbait tweets.  
For our analysis, we do not consider these users and their retweets, 
and only consider the users who exclusively participated in retweeting either clickbait or non-clickbait tweets in our dataset.

After gathering the tweets and retweets belonging to both clickbait and non-clickbait categories, in the subsequent sections, we attempt to answer the research questions we set out to investigate in this paper. 
\vspace{-3mm}
\section{RQ1. How are clickbait tweets different from non-clickbait tweets?}
\label{sec:RQ1}
In this section, we investigate how the clickbait and non-clickbait tweets differ from each other in terms of the tweet contents. 
Generally, we observe that clickbait tweets often mimic the article headlines, and both of them provide little 
information about the actual article contents. Here, the headlines and the tweets contain forward referencing 
cues to create a {\bf curiosity gap}~\cite{loewenstein1994psychology} to lure the readers so that they click on the article link to know more about the article. Whereas, non-clickbait tweets provide a summary of the content of the articles being referred to.

In the world of journalism, the process of the arousal of curiosity gap creates sensationalism. Just like clickbaits, any sensationalized content thrives on shock value. For this reason, tabloids in the past are known to use bold, attractive but sometimes misleading headlines. 
Such headlines contain something that the readers are already familiar with, and would want to know more~\cite{gans2009can}. 
In the subsequent subsections, we investigate the social media scenario -- whether the clickbait tweets, along with the arousal of curiosity, employ additional techniques to reach a wide user base.

\begin{figure}[t]
\centering
{
  \includegraphics[width=0.7\textwidth]{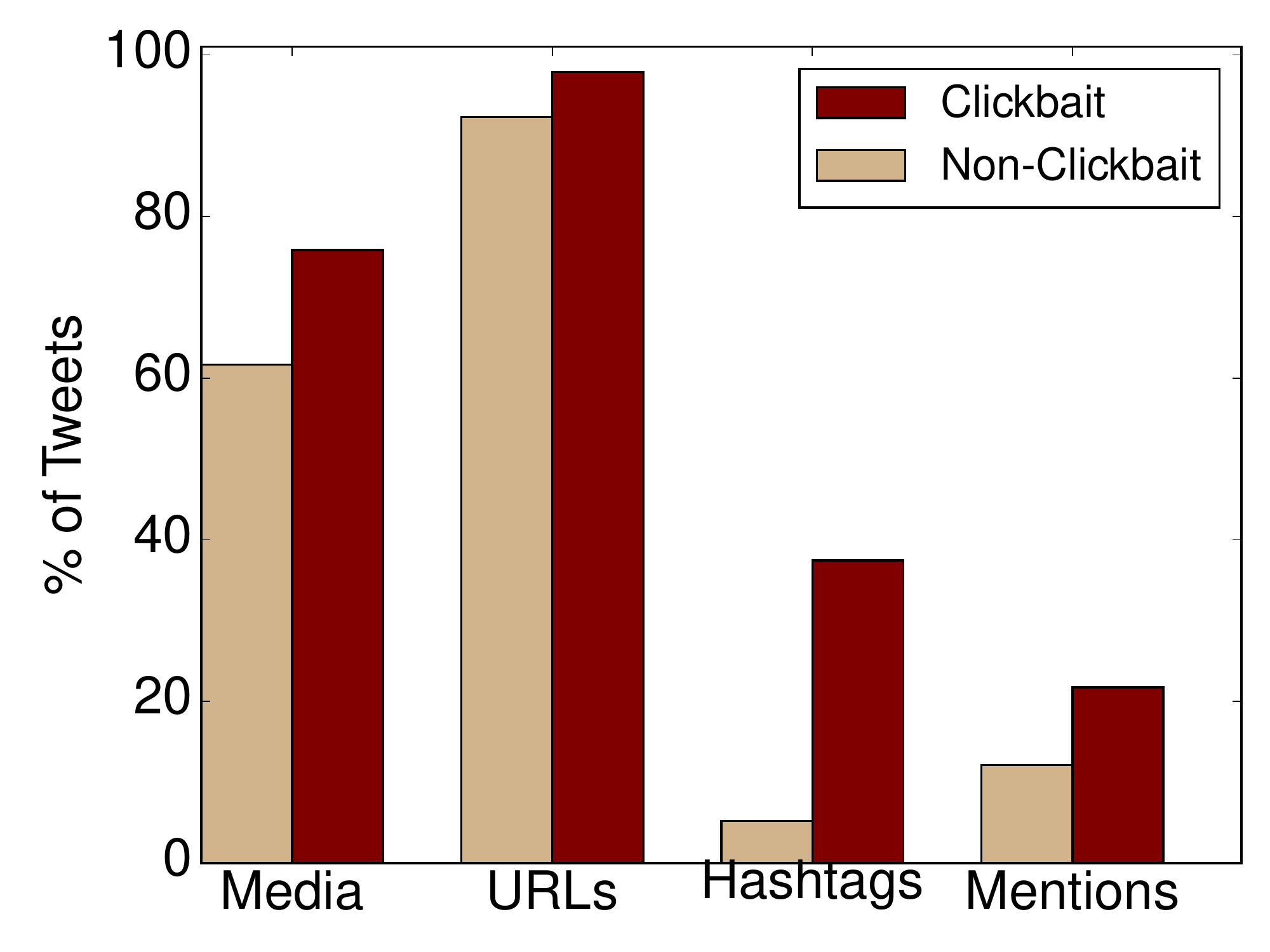}
  \caption[]{{\bf Comparing the presence of different entities in both clickbait and non-clickbait tweets. We conducted the analysis over tweets posted everyday during our 8 months data collection period, and present the average values here. All the differences are found to be statistically significant with $ p < 0.05$ in Welch's T-tests.}}
  \label{fig:tweet_content}
}
\vspace{-5mm}
\end{figure}

\subsection{Presence of different entities in the tweets}
Apart from the main textual content, a tweet also contains other entities such as images, videos, URLs, hashtags and mentions of other users. 
Including media content (in the form of images and videos) in the tweets help in increasing its attractiveness and amplifying the curiosity among the readers. 
Twitter provides an ubiquitous search box to search for any hashtag, and returns a list of tweets containing that particular hashtag. 
Therefore, including hashtags in the tweets help in reaching an wider audience on Twitter organically.

We start by comparing the presence of these different entities in both clickbait and non-clickbait tweets. Fig.~\ref{fig:tweet_content} 
shows the percentage of tweets in these two categories with at least one of the corresponding entities present, where the differences between the two categories are statistically significant ($ p < 0.05$ in Welch's T-tests~\cite{ruxton2006unequal}).
We can see in Fig.~\ref{fig:tweet_content} that the percentage of clickbait tweets containing media contents, user mentions, URLs  
and hashtags are significantly higher than that of non-clickbait tweets. 

While considering the average number of entities per tweet, we find that the clickbait tweets contain almost nine times the average number of hashtags ($0.37$) as compared to non-clickbait tweets ($0.041$). Similarly, clickbait tweets also contain a higher number of user-mention per tweet ($0.427$) compared to the non-clickbait tweets ($0.305$). Performing {\it Welch's T-tests}~\cite{ruxton2006unequal} over the entire corpus of clickbait and non-clickbait tweets confirm that the differences are {\it statistically significant} with $p < 0.05$. 

Increased use of media elements (such as images) in clickbait tweets can be compared with the tabloids, which brought in a sudden increase in the number of visual elements in newspaper design. Similar to clickbaits, tabloids have also been known to edit and enhance photographs to suit the tone of the news, and reserve a greater space for colorful graphic advertisements~\cite{randevnature}.

Similarly, increased use of user mentions, as well as the use of hashtags, not only lead to a greater organic reach, but also make a personal connect with the audience. A similar attempt can be witnessed in tabloids as well, when a public content is presented in a personal way, be it the story of a victim, or the private matters in the life of an individual (e.g., affairs, relationships, disagreements etc). By doing so, tabloids ensure that the news has special relevance for the isolated reader~\cite{ruberynewspaper}.

\subsection{Sentiment conveyed by the tweets}
Now, we attempt to analyze the sentiments conveyed through both non-clickbait and clickbait tweets. 
To determine the sentiments of the tweets, we used the sentiment analysis tool developed by Rouvier {\it et al.}~\cite{rouvier2016}, which uses an ensemble of Convolutional Neural Networks (CNN) with embeddings trained on different units: lexical, part-of-speech, and sentiment embeddings. In the final fusion step, the hidden layers of the CNNs are concatenated and fed into another deep neural network for sentiment detection~\cite{rouvier2016}. Given a tweet, it classifies the sentiment of the tweet into one of the three classes: positive, neutral, and negative. 

To provide a benchmark for sentiment analysis of tweets, ACL SIGLEX ({\tt siglex.org}) organizes {\it International Workshop on Semantic Evaluation (SemEval)}, which runs the competitive {\it Sentiment Analysis in Twitter} task every year~\cite{nakov2016semeval}. The tool developed in~\cite{rouvier2016} ranked 2nd at the SemEval task of 2016, achieving average F1 score of $63\%$ over three sentiment categories. In this work, we used the tool developed by Rouvier {\it et al.}~\cite{rouvier2016} primarily because of its superior performance, and also because the authors have kindly made the tool publicly available\footnote{available at \url{gitlab.lif.univ-mrs.fr/mickael.rouvier/SemEval2016}}. While analyzing the sentiment of clickbait and non-clickbait tweets, we first exclude the URLs associated with the tweets, and discard those tweets whose remaining lengths are less than three words.

We find that a higher fraction of non-clickbait tweets are associated with negative sentiment ($26.12\%$) as compared to clickbait tweets ($17.65\%$). Whereas, $40.75\%$ clickbait tweets convey positive sentiments compared to only $15.44\%$ non-clickbait tweets. This might be attributed to the fact that the news organizations tend to report on events which denote significant disruptions in the society, and often such news contain high negative sentiments. So, the proverb goes: \textit{No news is good news, and good news is no news.} On the contrary, clickbait tweets convey more positive sentiment  by covering mostly soft human interest stories to please their consumers.

Quite contrary to clickbaits, the offline media has been historically chided about the negativity and over sensationalization in the content of tabloids. In fact, Turner~\cite{turner1999tabloidization} acknowledged that the tabloid press prefers information over entertainment, or accuracy over sensation in their representations that exploit their audience. Often their focus is on staged family conflicts in an issue of public interest (e.g. a politician's private life). Interestingly, opposite to tabloids, clickbaits try to garner more attention by avoiding negative content.

\vspace{2mm}
\noindent \textbf{Summary} \\
In summary, we find that clickbait tweets contain significantly higher fraction of images, hashtags and user mentions compared to non-clickbait tweets, to engage with a broader set of audience. The tweet texts in clickbaits often offer little detail about the article being referred to, which leads to an increase in the curiosity of the consumer. As a result, clickbait tweets contain more external article links than non-clickbait tweets. Finally, clickbait tweets convey more positive sentiments than non-clickbait ones, in contrast to the offline tabloids which thrived by propagating more negative sentiments.
\section{RQ2. How do clickbait production and consumption differ from non-clickbaits?}
\label{sec:RQ2}
After studying the tweet contents, the next question we investigate is the difference between the production and consumption of clickbait and non-clickbait tweets. We define the {\it producers} of clickbait tweets as those users who have originally posted the tweets, and {\it consumers} as the users who have retweeted a clickbait tweet. A similar definition is established for the producers and consumers of non-clickbait tweets. We identify retweets by investigating  the `retweeted\_status' field in the tweet metadata returned by Twitter API. If a particular tweet is a retweet of another tweet, then this filed contains information about the original tweet that was retweeted. However, if some users copy and paste other tweets instead of retweeting them, then we can not track such duplicated tweets. 
We also acknowledge the limitation of using retweeting as a proxy for consumption. However, there is no direct way available to us to measure consumption at a large scale. 

To answer RQ2, we first study the diurnal variation in the activities of the producers and the consumers of both clickbait and non-clickbait tweets.

\begin{figure*}[tb]
\begin{minipage}[b]{0.493\textwidth}
\includegraphics[width=0.99\textwidth]{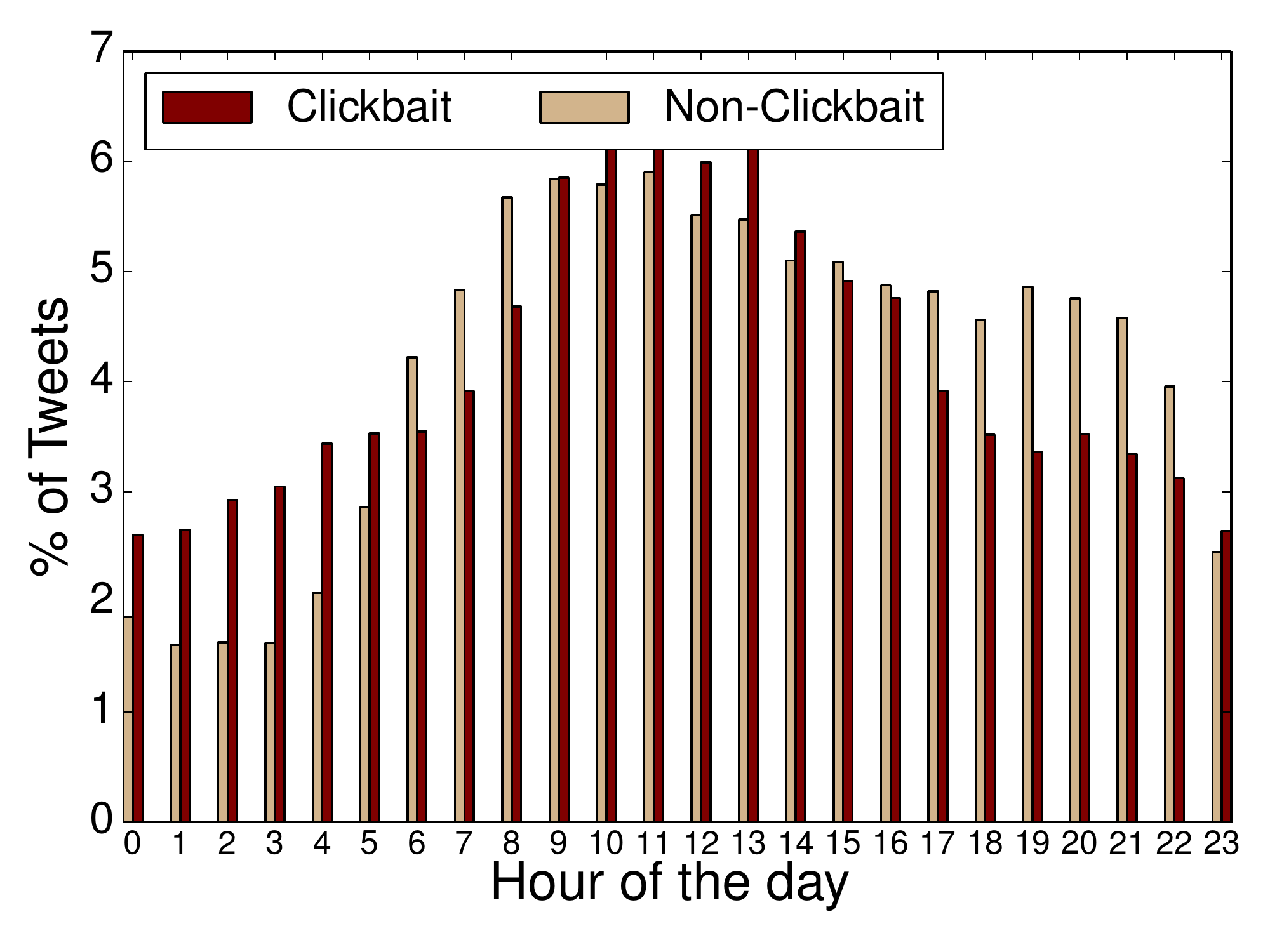}
\subcaption{}
\end{minipage}
\hfil
\begin{minipage}[b]{0.493\textwidth}
\includegraphics[width=0.99\textwidth]{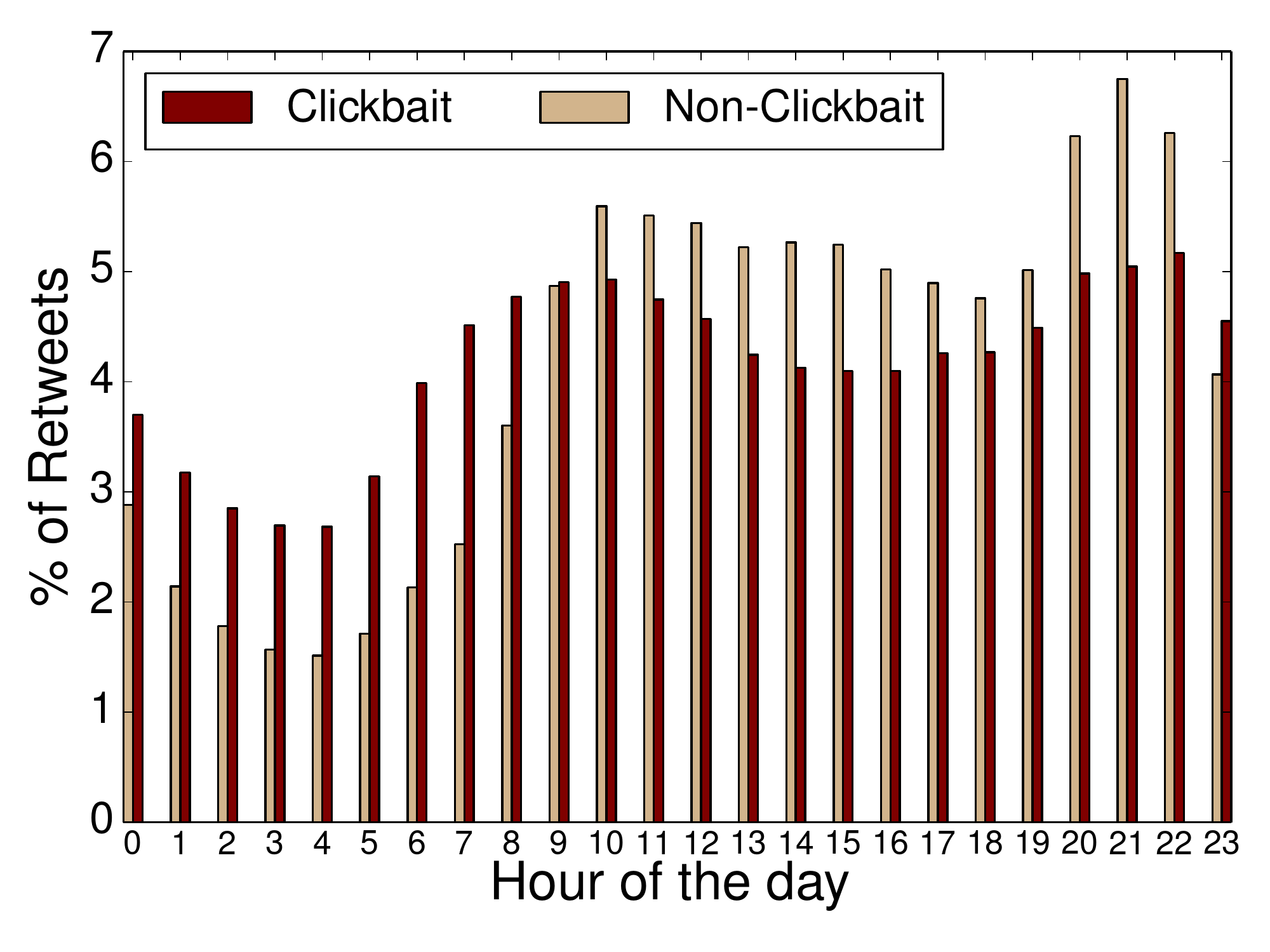}
\subcaption{}
\end{minipage}
\vspace*{-1mm}
\caption[]{{\bf Diurnal variations in (a) tweeting (i.e., production) and (b) retweeting (i.e., our proxy for consumption) of both clickbait and non-clickbait tweets posted everyday during our 8 months data collection period. Only except 9 AM, the differences in average values for all other hours are found to be statistically significant with $ p < 0.05$ in Welch's T-tests.}}
\label{fig:tweet_rt_time}
\vspace*{-3mm}
\end{figure*}

\subsection{Diurnal variations in tweeting and retweeting}
We attempt to observe the diurnal variations in the activities of both clickbait and non-clickbait producers 
(mostly comprising different media sources), i.e., we want to see how active they remain at different times of the day. 
This investigation would tell us whether a specific category of tweet (e.g., clickbait) is mostly posted during leisure hours.

For both clickbaits and non-clickbaits, different Twitter accounts may be operated from different timezones. Hence, we convert the posting times of every tweet to the corresponding local time of the user posting them (by using the timezone information of the user as returned by Twitter API). Similarly, as the consumers of these tweets may be distributed worldwide, we also convert the retweet times to the users' respective local times. The normalization helps in a better comparison between the activities across timezones.

Fig.~\ref{fig:tweet_rt_time}(a) shows the hourly variations in the production of both clickbait and non-clickbait tweets. Only except 9 AM, the fractions of clickbait and non-clickbait tweets being produced in all other hours are {\it significantly} different  ($p < 0.05$ in the corresponding Welch's T-tests). Similarly, Fig.~\ref{fig:tweet_rt_time}(b) shows the hourly distributions of retweeting activity in both clickbait and non-clickbait tweets. Even in case of retweets, we observe significant difference in hourly activities of clickbait and non-clickbait consumers.

We see some common features between the overall patterns in both production and consumption activity in Fig.~\ref{fig:tweet_rt_time}(a), and in Fig.~\ref{fig:tweet_rt_time}(b). For instance, both production and consumption is relatively less during the night (10 PM to 5 AM) for both clickbaits and non-clickbaits compared to their day activities. However, we also notice some interesting differences in the production and consumption activities of the two categories. For instance, clickbait tweets tend to be posted more during peak hours of the day (from 10 AM to 2 PM) as well as on night (from 11 PM to 5 AM). Whereas, consumption of clickbait tweets is higher in the odd hours of the night and early morning (between 11 PM and 8 AM), but for the rest of the day, users tend to consume more non-clickbait tweets.

\subsection{Longevity of individual tweets}
\begin{figure*}[tb]
\begin{minipage}[b]{0.493\textwidth}
\includegraphics[width=0.99\textwidth]{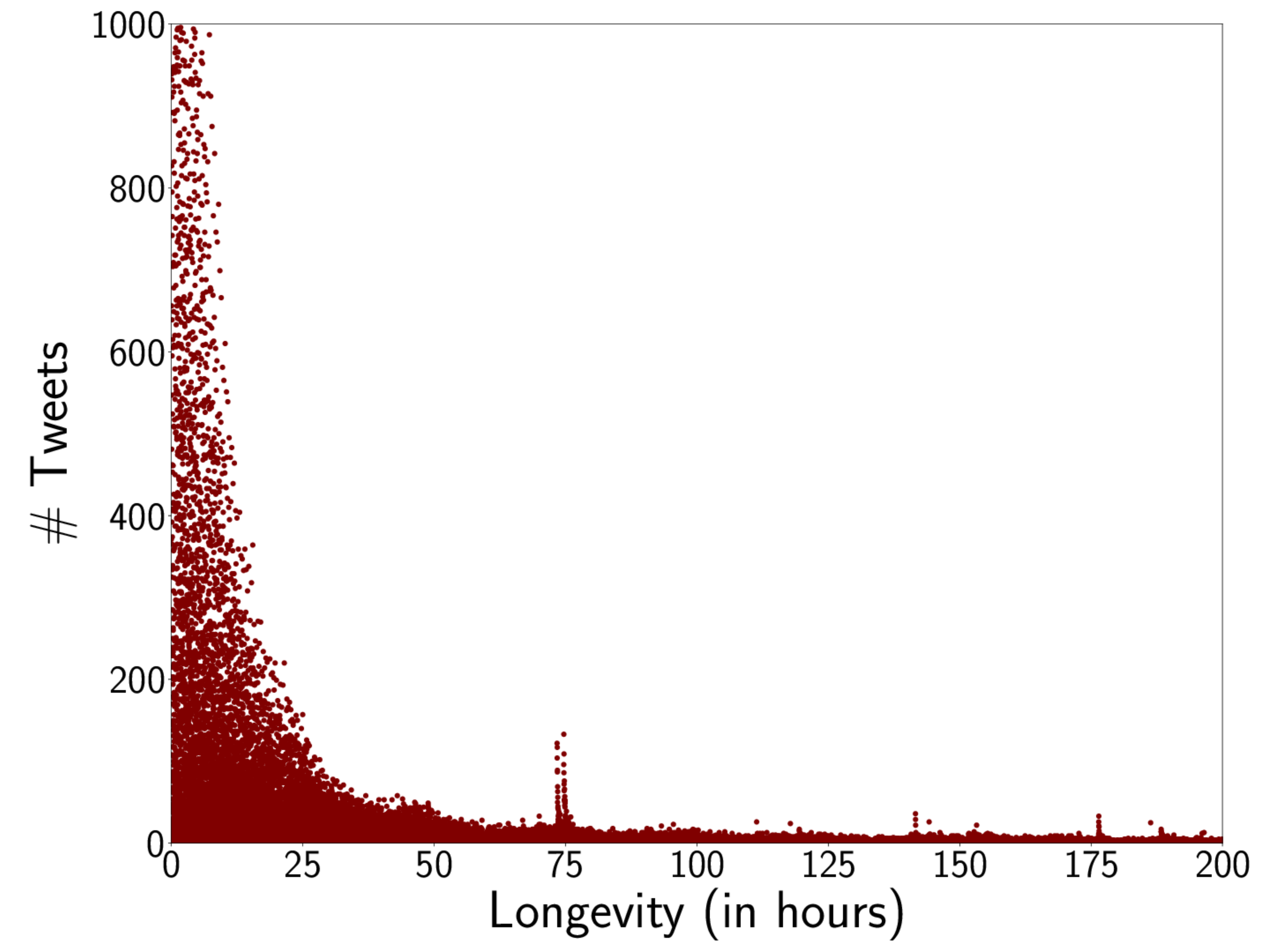}
\subcaption{}
\end{minipage}
\hfil
\begin{minipage}[b]{0.493\textwidth}
\includegraphics[width=0.99\textwidth]{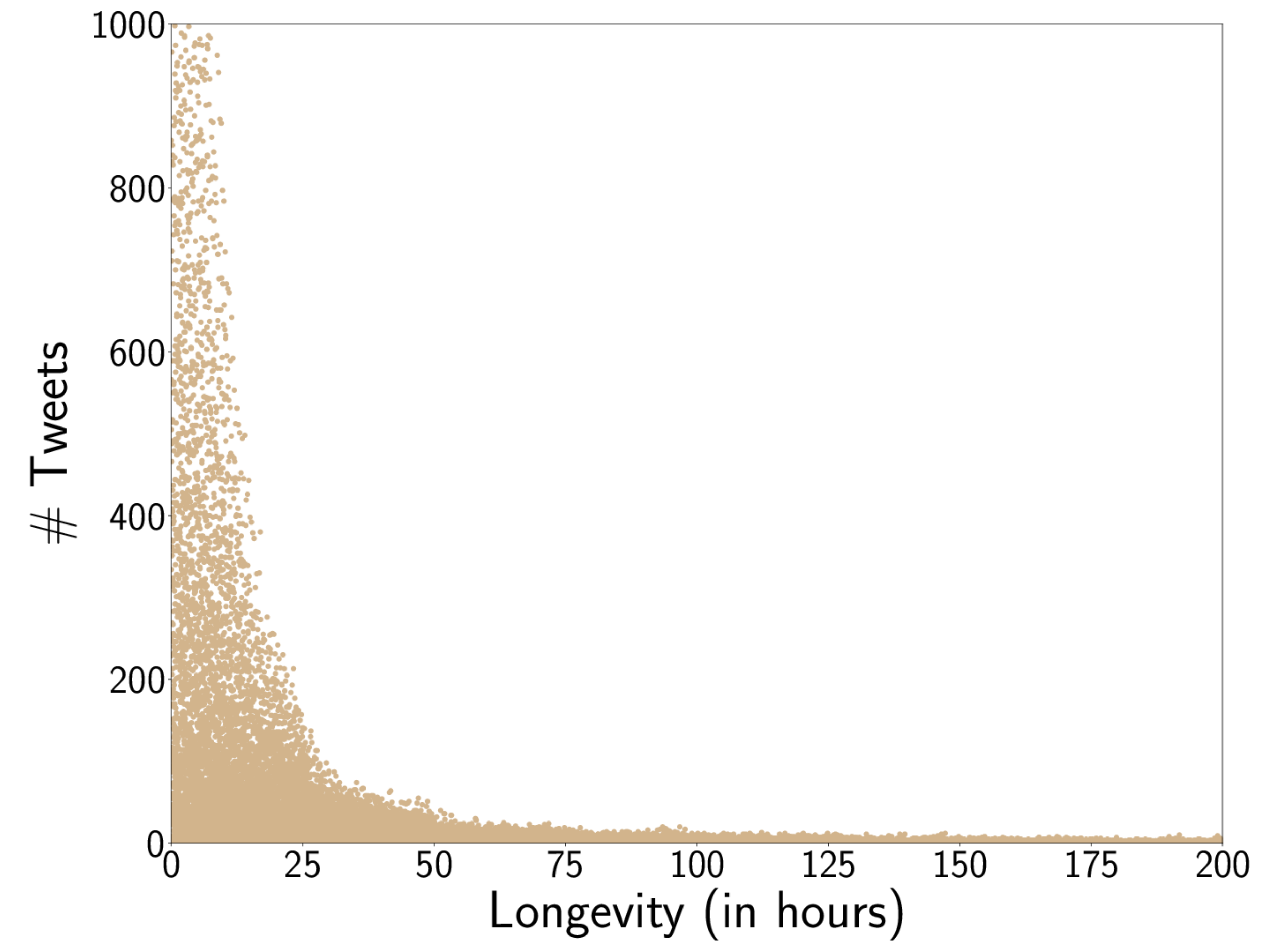}
\subcaption{}
\end{minipage}
\vspace*{-1mm}
\caption[]{{\bf Scatter plot of longevity values (in hours) for different tweets in (a) Clickbait, and (b) Non-clickbait category.}}
\label{fig:longevity}
\vspace*{-3mm}
\end{figure*}

Next, we attempt to measure how long the consumers remain engaged with individual clickbait and non-clickbait tweets. 
Towards that end, we chronologically order the retweets for a tweet, and we define the {\it longevity of a tweet} as the 
difference between the time when the original tweet was posted and the time of the $95$th percentile retweet of 
the original tweet (in our dataset). Fig.~\ref{fig:longevity}(a) and Fig.~\ref{fig:longevity}(b) show 
the longevity of different tweets in both clickbait and non-clickbait categories. We can see clear distinction between the engagement levels 
in clickbait and non-clickbait tweets. We find that clickbait tweets have higher user engagement levels 
as compared to that of non-clickbait tweets. The spikes in the longevity of clickbait tweets show that higher number of 
clickbait tweets received retweets beyond $50$ hours from the posting of the original tweet. 

Such difference in engagements possibly arise because the non-clickbait tweets tend to focus on some temporal 
events (e.g., political rallies, sports matches, entertainment events etc.), and hence the user attention decays 
rapidly with time as the relevance of the corresponding event fades away. On the other hand, there is less time binding associated with 
clickbait tweets, and therefore, they tend to attract user engagements beyond any specific time threshold.

\subsection{Length of the retweet cascade}
One of the main characteristics of social networks like Twitter is that it allows its users to post and {\it reshare} content they think would be 
useful for their followers. In some cases, an original tweet gets re-(and-re-re)tweeted multiple times. For example, a user 
first posts a tweet and then one of her followers shares the tweet with her set of followers, and some of these followers retweet to 
their respective sets of followers, and so on. In this process, a {\it retweet cascade}~\cite{cheng2014can} gets formed, 
thereby potentially reaching a large audience. Information cascades have been studied in multiple different settings such as {\it viral marketing}~\cite{leskovec2007dynamics} or email-chains~\cite{liben2008tracing}, and it has been noted that the larger the cascades are, higher reach they have in terms of audience size. 

After observing that clickbait tweets tend to get user attention for longer duration, 
a related question to ask is whether the length of the retweet cascades formed in clickbait and non-clickbait tweets 
are similar. We calculate the cascade lengths of tweets belonging to both clickbait and non-clickbait categories, and find that most of the 
tweets in both categories have cascade length $1$, denoting the retweets made within the direct followers. This finding corroborates with earlier studies which observed that large cascades are rare~\cite{leskovec2007dynamics,liben2008tracing}. However, on considering the 
$90$-th percentile cascade lengths, we see that the clickbait tweets (with the value of $6.47$) could penetrate the users to a deeper 
extent, and hence have a higher reach than the non-clickbait tweets (with the $90$-th percentile cascade length being $5.23$). 

\vspace{2mm}
\noindent \textbf{Summary}\\
By conducting the above analyses, we see that the clickbait tweets retain their popularity among their user base much longer as compared to non-clickbait tweets. The followers of users retweeting clickbait tweets take active notice and subsequently participate in its propagation. This can largely be attributed to the fact that clickbait tweets are generally associated with topics which are not time-bound, and they seem more attractive to the consumers by virtue of the arousal of curiosity gap, as described earlier.
\section{RQ3. Who are the consumers of clickbait and non-clickbait tweets?}
\label{sec:RQ3}
As mentioned in Section~\ref{sec:related}, prior readership surveys have found that the traditional broadsheet newspapers mostly cater to the affluent and well educated audience. Whereas, the majority of the tabloid readers are found to be in the lower middle and working class. Following that line, in this section, we try to understand the distinction between the users who consume clickbait and non-clickbait tweets. We compare and contrast their popularity and reputation on social media (which can act as a proxy to their class information), and their demographics. By analyzing the popularity and reputation, we hope to conclude whether a type of tweet is consumed by more popular users (users having high follower count) or by less popular users. Also, by studying the demographics of consumers such as their gender and age, we expect to get an account of what section of the society follows clickbait or non-clickbait tweets. 

\begin{figure*}[tb]
\begin{minipage}[b]{0.493\textwidth}
\includegraphics[width=0.99\textwidth]{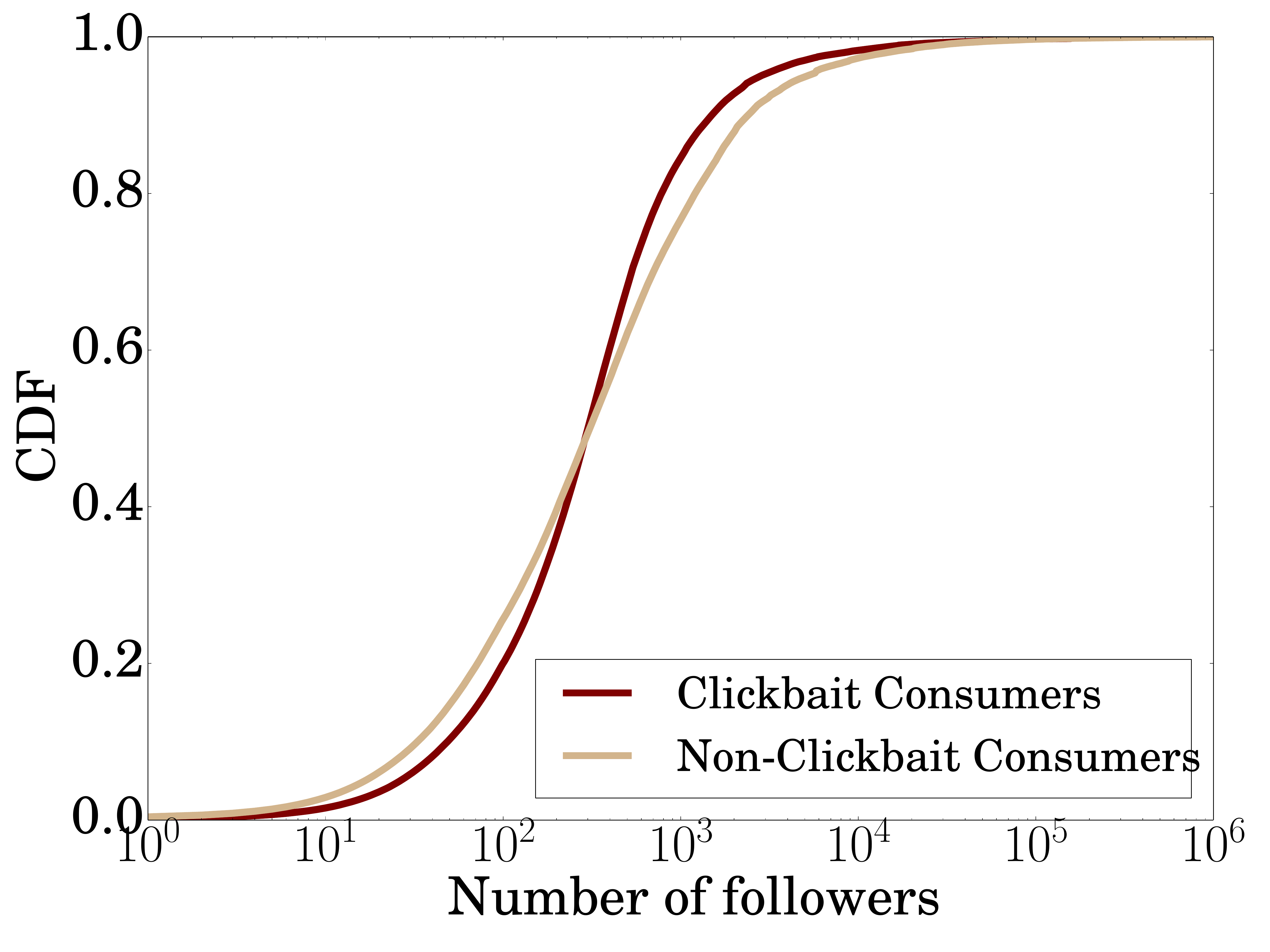}
\subcaption{}
\end{minipage}
\hfil
\begin{minipage}[b]{0.493\textwidth}
\includegraphics[width=0.99\textwidth]{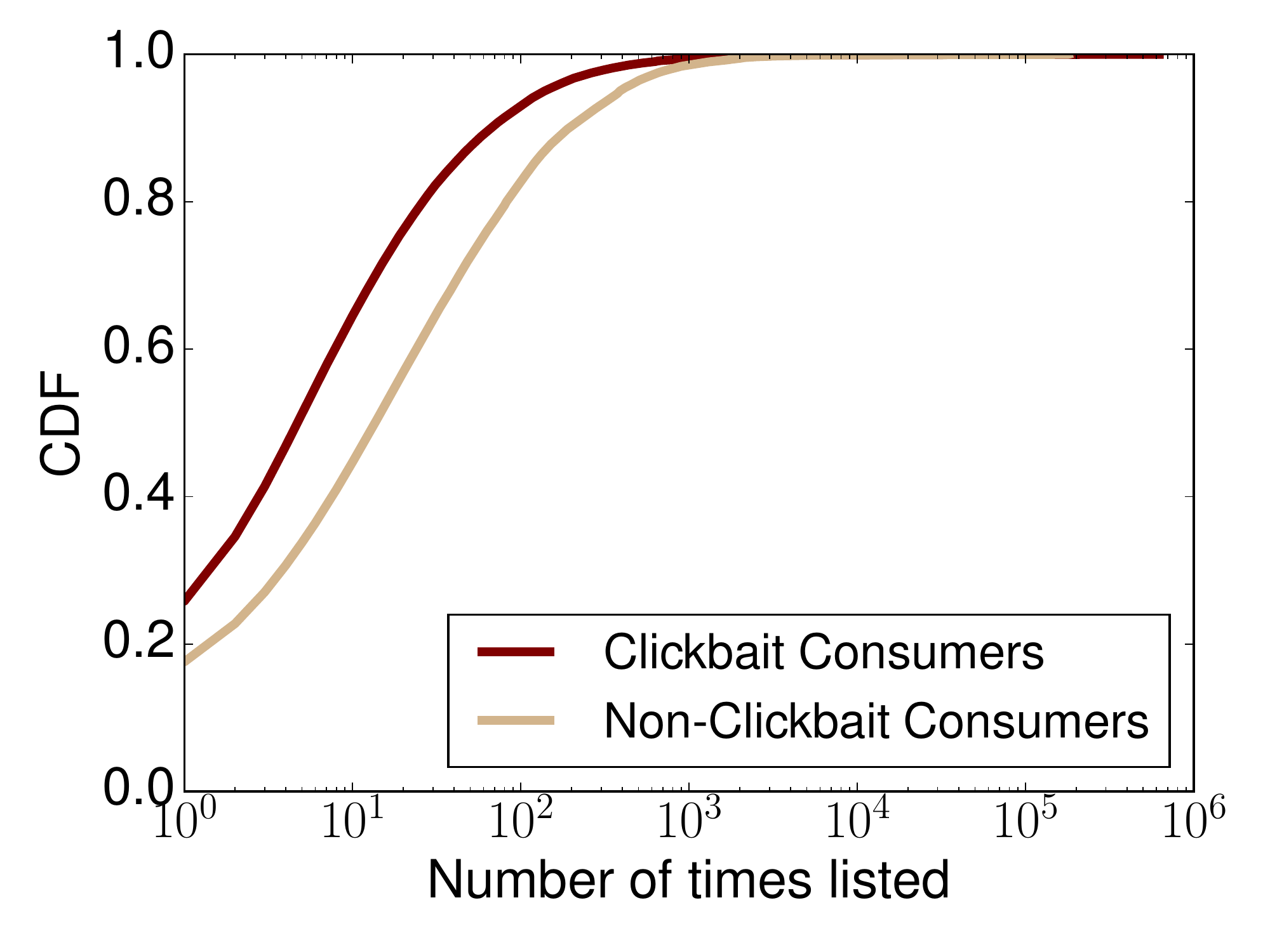}
\subcaption{}
\end{minipage}
\vspace*{-1mm}
\caption[]{{\bf For both clickbait and non-clickbait consumers, CDF of (a) number of their followers, and (b) number of times they are listed.}}
\label{fig:follower_listed_count}
\vspace*{-3mm}
\end{figure*}

\subsection{Popularity and reputation of consumers}
Next, we turn our focus towards understanding how popular the consumers of clickbait tweets are vis-a-vis the consumers of non-clickbait tweets. Depending on the number of followers they have, we group the Twitter users into three different categories: \\
(i) Less popular users (having less than $ 1,000$ followers), \\ 
(ii) Popular users (having $1,000$ to $10,000$ followers), and \\
(iii) Very popular users (having more than $ 10,000$ followers).

Fig.~\ref{fig:follower_listed_count}(a) shows the cumulative distribution function (CDF) of the number of followers 
for both clickbait and non-clickbait consumers.
We can see in Fig.~\ref{fig:follower_listed_count}(a) that among the less popular range, 
on average, clickbait consumers have more followers than non-clickbait consumers. In other words, among the less popular users, 
clickbait consumers tend to be followed more than non-clickbait consumers.
However, this trend reverses among popular users, where non-clickbait consumers are more followed than clickbait consumers. 
Finally, we don't find much difference in number of followers of the very popular users for both clickbait and non-clickbait tweets.

In addition to comparing the number of followers, we also look at the number of lists in which the consumers are enlisted.  
{\it Lists} is a feature in Twitter, using which a user can create a list of other users who tweet on a particular topic. 
For example, a user may create a list `Politics' to include @realDonaldTrump, @BarackObama, @whitehouse, or @HillaryClinton.
Similarly, another list `Music' may include @britneyspears, @ladygaga, or @rihanna. 
The Twitter users present on these lists can be thought as the {\it experts} on that topic. 
More a particular user is enlisted in different lists, she can be regarded as more reputed in the Twitter community. 
There have been prior research works which have used these lists to build expert search and recommendation systems~\cite{ghosh2012cognos,sharma2012inferring}. 

In our context, we use this list feature to compare the reputation of both clickbait and non-clickbait consumers. 
More specifically, we compare the number of times different consumers are enlisted in different lists. 
As can be seen in the CDF plot in Fig.~\ref{fig:follower_listed_count}(b), non-clickbait consumers are listed much more on average than the clickbait consumers. Therefore, we can conclude that non-clickbait consumers are more reputed in the community than their clickbait counterparts.

\subsection{What the profile descriptions reveal}
Next, we analyze the self-identified Twitter biographies of the consumers. 
We perform an LIWC~\cite{tausczik2010psychological} analysis on the words used by the clickbait and non-clickbait consumers in their profile bios. We notice a stark difference in the use of several linguistic categories by clickbait consumers and their non-clickbait counterparts (the differences are found to be {\it statistically significant} in Welch's T-tests~\cite{ruxton2006unequal}). For example, clickbait consumers tend to use swear words $130.77\%$ more often than non-clickbait consumers ($0.3$ vs $0.13, p <0.05$). 
Additionally, the use of personal pronouns by clickbait consumers is $30.04\%$ more than non-clickbait consumers ($6.71$ vs $5.16, p <0.05$). 
We also note a significant jump in the use of words with negative tones by clickbait consumers. The use of anxiety words like nervous, afraid and tense is $95.24\%$ more by clickbait consumers than by their non-clickbait counterparts ($0.41$ vs $0.21, p <0.05$). They are also $63.16\%$ more frequent in using words like grief, cry and sad ($0.31$ vs $0.19, p <0.05$). Clickbait consumers use $80\%$ more filler words like `you know' and `I mean' ($0.18$ vs $0.1, p <0.05$). 

Non-clickbait consumers, on the other hand, are $27\%$ more likely to use family related words ($1.07$ vs $0.84, p <0.05$), 
and are $56\%$ more frequent in using words related to work, achievements and their occupational roles (e.g.,~\textit{lawyer},~\textit{scientist}, ~\textit{reporter}) ($5.93$ vs $3.8, p <0.05$). Moreover, their interests seem to be in the areas of business and economic affairs, as reflected by their $65\%$ higher use of words like audit, cash and owe (pertaining to money) ($1.09$ vs $0.66, p <0.05$).  
We also notice words like~\textit{girl} and~\textit{princess} appearing more among the bio of clickbait consumers; while the word~\textit{father} appearing more in its non-clickbait counterpart. This indicates a tilting of women towards clickbait articles and more male prominence among the non-clickbait consumers. We now look at the demographic distribution of the consumers to more thoroughly investigate this tilt.

\subsection{Consumer demographics}
To compare the demographics of the clickbait and non-clickbait consumers, we consider their gender, and age. 
However, inferring the demographics of Twitter users at scale is challenging. There have been some approaches for inferring the gender of a Twitter user from the username~\cite{blevins2015jane}, 
or the age from Twitter profile description (by finding textual patterns like `21 yr old', `born in 1989')~\cite{sloan2015tweets}.
However, due to their reliance on standard usernames and profile descriptions, these approaches fail to infer the demographics for a large number of users~\cite{chakraborty2017makes}. 
Following the approach taken by some of the earlier attempts in working with user demographics
~\cite{Jisun2016,chakraborty2017makes}, 
we use {\it Face++ API}~\cite{facepp}, 
a face recognition platform based on deep learning~\cite{yin2015learning},
to extract the gender and age from the recognized
faces in the profile images of the Twitter users. Some profile images may not have any 
recognizable face, while some other images may have more than one faces. 
Discarding these images (and the image URLs which were unavailable at the time of the inference), 
we end up getting the demographic information for around $208K$ clickbait consumers 
and $226K$ non-clickbait consumers. Face++ returns `Male' or `Female' as the gender, 
and a numerical value as the age of a user. 

\begin{figure*}[tb]
\begin{minipage}[b]{0.493\textwidth}
\includegraphics[width=0.99\textwidth]{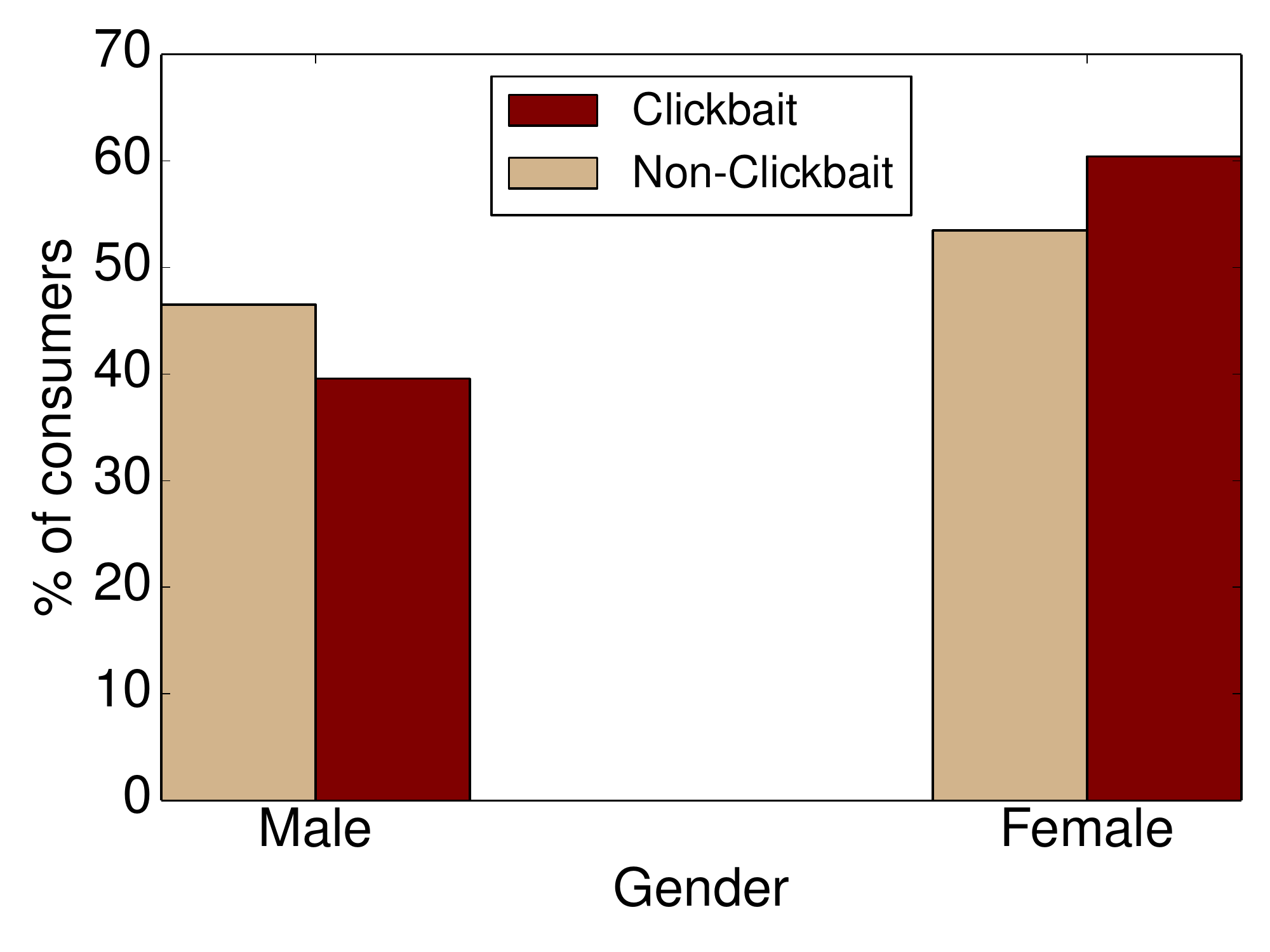}
\subcaption{}
\end{minipage}
\hfil
\begin{minipage}[b]{0.493\textwidth}
\includegraphics[width=0.99\textwidth]{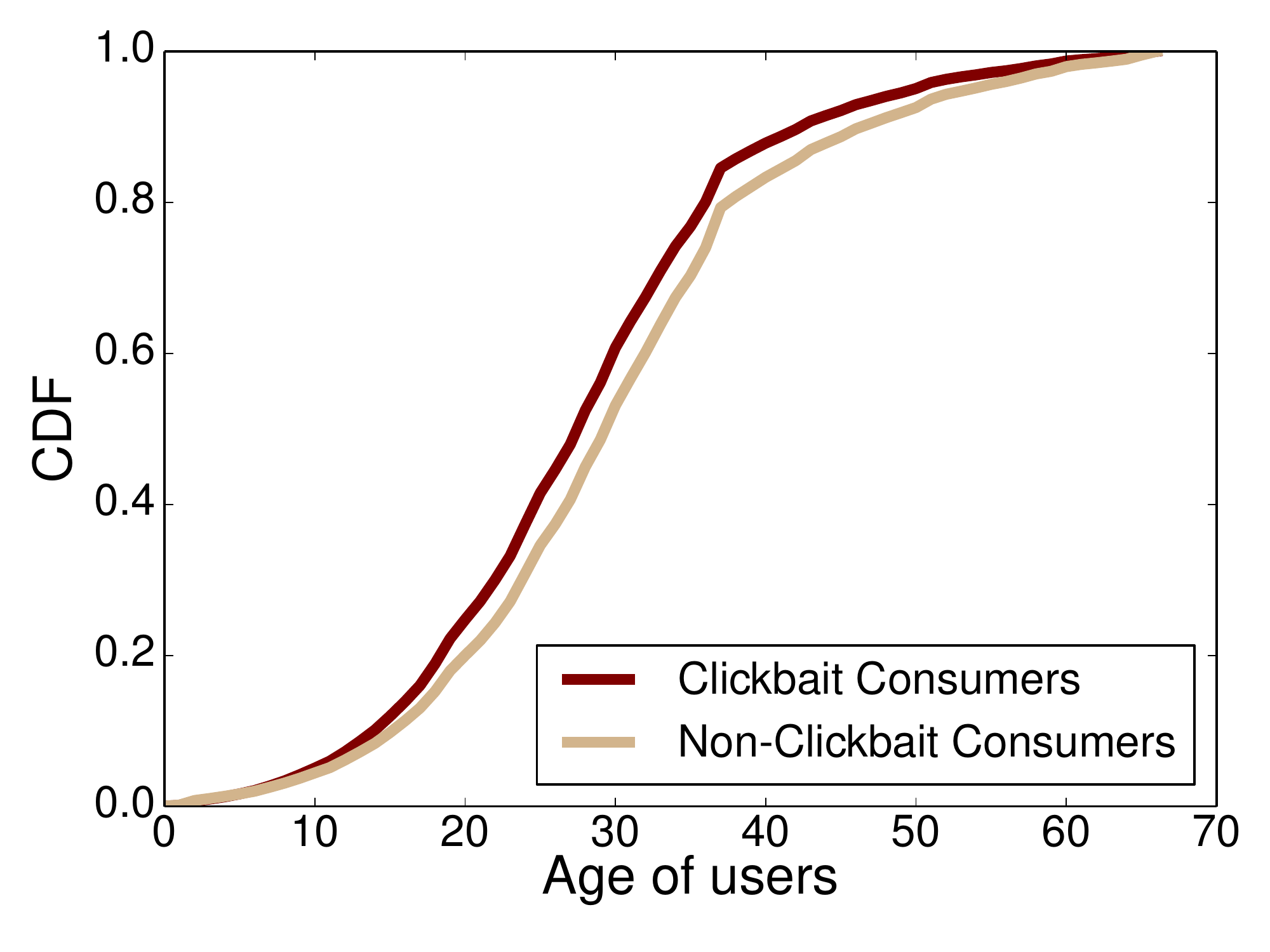}
\subcaption{}
\end{minipage}
\vspace*{-1mm}
\caption[]{{\bf Demographics of clickbait and non-clickbait consumers: (a) their gender distribution, and (b) CDF of their age.}}
\label{fig:demographics}
\vspace*{-5mm}
\end{figure*}

To evaluate the performance of Face++, we first collected a set of users for whom we can 
infer the gender and age from their screen names and profile descriptions. 
Among the users for whom we got the inferred demographics from Face++, we could successfully get the gender from the screen name (using the method developed by Blevins {\it et al.}~\cite{blevins2015jane}), and age from the profile description (using the method proposed by Sloan {\it et al.}~\cite{sloan2015tweets}) for $3,208$ users. 
Then, we computed the accuracy of the demographic inference by Face++, by considering the data obtained for these $3,208$ users as ground truth.
We found the gender inference accuracy of Face++ to be $84\%$; whereas, the average error in the age inferred by Face++ was $5.2$ years. 
We note that the accuracy results are similar to the ones reported in earlier evaluations~\cite{Jisun2016,chakraborty2017makes}.

Using the demographic information returned by Face++ for both clickbait and non-clickbait consumers, we 
compute their overall demographics considering their gender and age. 
Fig.~\ref{fig:demographics}(a) shows the gender 
distribution among both the clickbait and  non-clickbait consumers.
Fig.~\ref{fig:demographics}(b) shows the CDF of ages of the consumers of both categories. 

In a recent work, Chakraborty {\it et al.}~\cite{chakraborty2017makes} reported that among the Twitter users, 
there are more women than men ($53.1\%$ vs. $46.9$). Therefore, we can see in Fig.~\ref{fig:demographics}(a) that 
for both clickbait and non-clickbait consumers, there are more women than men. 
However, by comparing the fraction of women among the clickbait and non-clickbait consumers, 
we see that there are higher fraction of women among clickbait consumers than non-clickbait consumers. 
Whereas, it is the opposite when comparing the fraction of men among the consumers. 

Surprisingly, this observation is contrary to the gender distribution observed among tabloid readers, 
where the majority of readers are men~\cite{johansson2007reading}. Journalism researchers have argued that even though most of the tabloids provide woman's pages reporting on fashion and woman's health issues in addition to the celebrity gossip pages, sexualised representation of women in the overall news discourse attracts way more male readers to tabloids~\cite{johansson2007reading}.

Regarding the age of the consumers, Fig.~\ref{fig:demographics}(b) shows that the clickbait consumers 
tend to be younger than the non-clickbait consumers. Similar to clickbaits, a large fraction of tabloid readers are aged below 35 years, 
compared to much less fraction of readers aged above 65 years~\cite{johansson2007reading}.

\vspace{1mm}
\noindent \textbf{Summary} \\
By conducting the analyses in this section, we find that the clickbait tweets are more popular among 
women, 
and also among relatively younger Twitter users. 
On the other hand, non-clickbait consumers have a higher proportion of men and older people. 
Additionally, they are more reputed, and have relatively higher follower base. 
Profile descriptions of both type of consumers reveal significant differences in the use of words from different 
linguistic categories. 
\vspace{-2mm}
\section{RQ4. How do the clickbait and non-clickbait consumers differ as a group?}
\label{sec:RQ4}
Criticisms around tabloidization are mostly grounded in Habermas' notion of the {\it public sphere}~\cite{habermas1991structural}, where public opinion can be formed via {\it rational-critical debates} between private individuals. Although news media is considered to be the enabler of such communications, critiques of tabloids have argued that tabloids fail those standards to enable debates in the public sphere~\cite{johansson2007reading}.

In this work, we investigate a related question in the context of social media -- among clickbaits or non-clickbaits, which is a better enabler of  communication between different groups of people in the public sphere (i.e., the social media platform). To answer this question, we study the reciprocity of the follower graph, retweet graph, and mention graph constructed from the activities of both clickbait and non-clickbait consumers, where reciprocity determines the extent of mutual engagement between the consumers. We also analyzed the density of the graphs, but the graphs turn out to be very sparse for both type of consumers.

\subsection{Properties of the follower graph}
We can visualize the full Twitter network as a directed graph where every user has a link to her follower. Reciprocity in this graph is the fraction of user pairs having links both ways (i.e., they have bidirectional following relation). 
The follower graph of clickbait and non-clickbait consumers are two induced subgraphs from the large Twitter follower network.
We calculated the reciprocity in both subgraphs. However, we did not record significant difference in the results obtained for the two kinds of consumers. 
For the clickbait consumers, the reciprocity value is $0.00187$, while it is $0.00186$ for the 
non-clicbait consumers. The low reciprocity suggests that forging mutual following is not very common for both clickbait and non-clickbait consumers.

\subsection{Properties of the retweet graph}
We next analyze the retweet graph of clickbait and non-clickbait tweets. Two users in a retweet graph are connected by bidirectional edge when both have retweeted some tweets posted by the other. 
We find that the reciprocity in case of clickbait consumers ($9.74 e^{-4}$) is almost $1.5$ times that of the non-clickbait consumers ($6.04 e^{-4}$). The results tell us that the clickbait consumers tend to mutually engage more in terms of retweeting each other than non-clickbait consumers.

\subsection{Properties of the mention graph}
Similar to the retweet graph, in the mention graph, bidirectional edge is created when two users mention each other in their tweets. We next analyze the mention graph of clickbait and non-clickbait consumers. 
We find that the reciprocity for clickbait consumers ($1.63 e^{-3}$) is almost $2$ times that of the non-clickbait consumers ($8.17 e^{-4}$).
Therefore, similar to the retweet activity, clickbait consumers also tend to mention each other way more than their non-clickbait counterparts.

\begin{figure}[t]
\centering
{
  \includegraphics[width=0.6\textwidth]{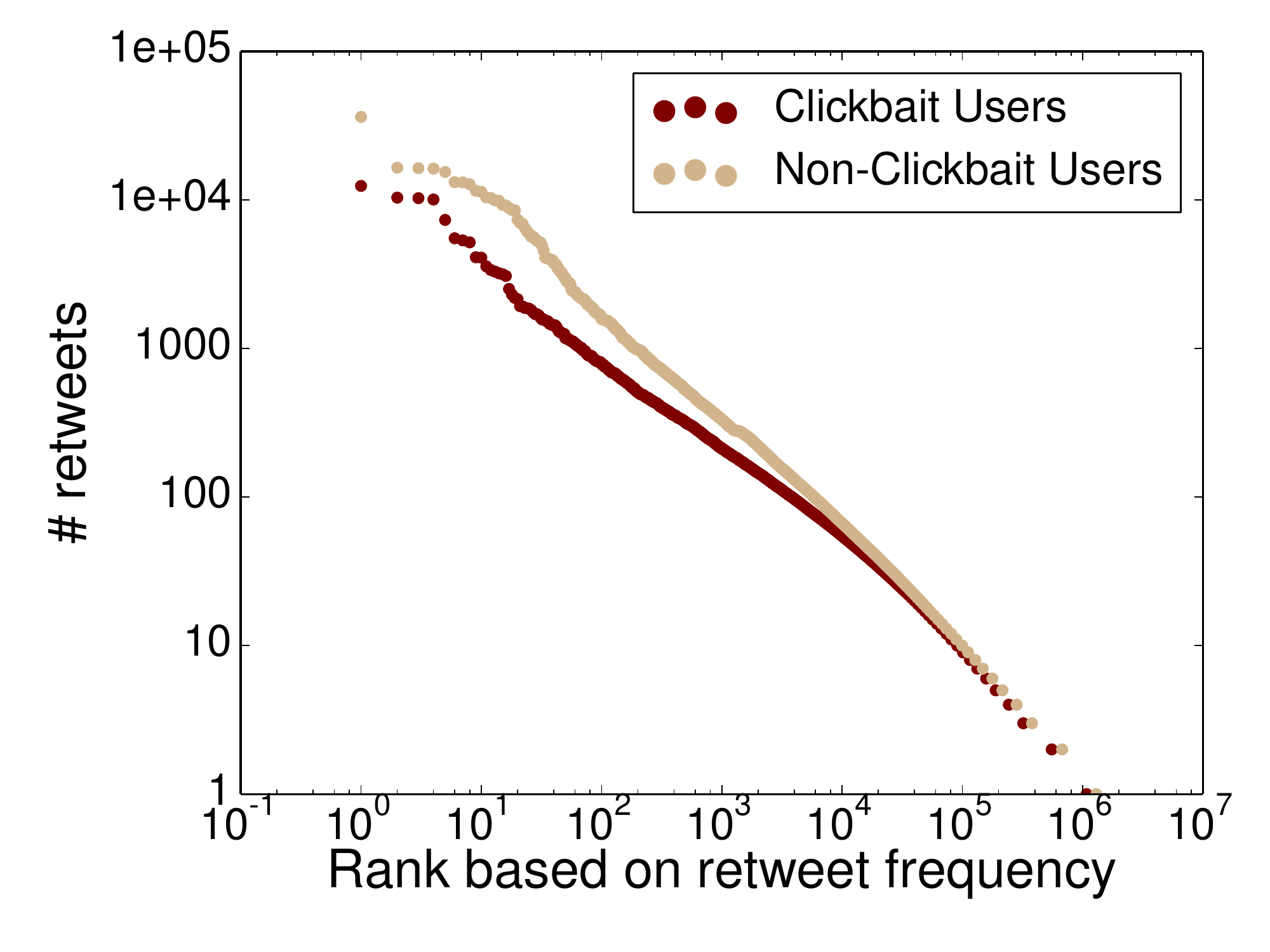}
  \vspace{-1mm}
  \caption[]{{\bf Retweet frequencies of clickbait and non-clickbait consumers.}}
  \label{fig:retweet_frequency}
}
\vspace{-0.5cm}
\end{figure}

\subsection{Retweet frequency of the users}
We now analyze the retweeting activities of both clickbait and non-clickbait consumers. 
To this end, we first rank the consumers based on the number of retweets they made, 
and then plot the ranks against the corresponding retweet count in Fig.~\ref{fig:retweet_frequency}. 
We observe in Fig.~\ref{fig:retweet_frequency} that the retweeting activities of both clickbait and non-clickbait 
consumers exhibit {\it heavy tailed} distributions where a small number of users have made most of the retweets. 

However, the distribution for clickbait consumers are relatively less skewed than that of non-clickbait consumers. 
This is interesting because when we compute the average number of retweets 
made, we find similar values for both clickbait ($5.257$), and non-clickbait consumers ($5.165$).
Such higher skew in non-clickbait consumers signifies the potential presence of a core group of consumers 
who are more active in retweeting non-clickbait tweets. Members of such core groups can play  
the roles of {\it opinion leaders} in the public sphere~\cite{richins1988role}, 
and initiate discussions in their communities around the non-clickbait stories. 
On the other hand, clickbait consumers have relatively more uniform retweeting activities 
implying less reliance on opinion leaders and more democratization of the discussions around clickbait stories.

\vspace{2mm}
\noindent \textbf{Summary} \\
Doing the above analyses, we find that the level of mutual engagement between clickbait consumers is more than that of non-clickbait consumers. Also, we find that clickbait consumers follow a much uniform pattern while retweeting as compared to the non-clickbait consumers.
For long, there have been many criticisms of Habermas' notion of public sphere. For example, Fraser argued that Habermas' conception is elitist, and therefore, members of subordinated groups require alternative arenas for {\it public discourse} to articulate their interests~\cite{fraser1990rethinking}. The idea of {\it alternative public spheres} has been extensively used to justify the effectiveness of tabloids in providing such arena~\cite{johansson2007reading,ornebring2004tabloid}. 
Likewise, the observations in this section highlight that in social media, clickbaits successfully provide such alternative public sphere for users drifting away from traditional news.
\vspace{-2mm}
\section{Concluding Discussion}
\label{sec:conclusion}
In this work, we analyzed the production and consumption of clickbaits in social media. We posed four related research questions 
which we answered using several analyses. We observed that there is a clear distinction in the way clickbait tweets and their consumers differ from that of non-clickbaits. 

Our investigation reveals several interesting insights. Clickbait tweets include more entities such as images, hashtags, and user mentions 
which help in capturing the attention of the consumers.  
It is also noted that a higher percentage of clickbait tweets convey positive sentiments as compared to non-clickbait tweets. 
Clickbait tweets tend to have a wider and deeper reach in its consumer base as compared to non-clickbait tweets. 

Additionally, we made interesting observations regarding the clickbait consumers. For example, clickbait tweets are consumed by more  women compared to their fraction among the consumers of non-clickbait tweets. 
Also, the clickbait consumers are younger than non-clickbait consumers. 
The results also point out that the non-clickbait consumers are more reputed in the community, and the linguistic composition of their profile descriptions differ significantly from the clickbait consumers. It can also be concluded the clickbait consumers have more mutual engagement among each other, and they retweet more uniformly than non-clickbait consumers.

The conclusion that the clickbait tweets are more popular among a section of the society which is clearly distinct 
from the consumers of regular news, again corroborates the general notion that the mainstream journalism is more for 
a certain class of the society (e.g., affluent and higher-educated). 
For a long time in history, the narrative in newspapers was also controlled by that class (including politicians, filmstars, sportsmen, large business owners, administrators and general experts). 
In this context, the tabloids emerged as a useful tool for raising the societal awareness among the {\it subalterns}~\cite{baum2006oprah}. 
We can hypothesize that just like tabloids, clickbaits also cater to the section of the society which shows little interest in following mainstream journalism.

In fact, it is noted that the criticism of tabloid journalism and tabloid form in general is more often made using traditional criteria of political power (voting, participation in formal political activities etc.), rather than the criteria of cultural recognition (representation, participation in alternate political forms etc.)~\cite{ornebring2004tabloid}. 
Similarly, prior works on clickbaits report only the negative aspects of it, by turning a blind eye to the vast majority of users who are embracing clickbaits. We can extend the earlier argument to point out that only criticizing clickbaits and trying to get rid of them altogether may not be desirable. 
Instead of blanket prevention of clickbaits, 
we believe that a four-pronged approach should be taken to tackle the prevalence of clickbaits.

First, an alternate solution may be to indicate the {\it newsworthiness} of an article to the intended readers. It can be integrated in form of the browser extensions developed in~\cite{chakraborty2016stop,downworthy}. When a user hover the mouse over an article link, she can be shown a short summary of the article being referred to, along with the possible news value of the article. Such an option would bring in more transparency to the users, and they can decide whether to pursue a particular article or not. 

Second, if certain groups of users indeed pursue clickbaits more, dedicated clickbait recommender systems can be developed to deliver more values to them. In such systems, relatively better quality articles may be given higher importance, which can potentially weed out lower quality information and encourage clickbait producers to come up with attractive yet informative article content.

Third, clicks are important in today's advertisement driven revenue model. Hence, motivated by the success of clickbait headlines, many traditional media organizations have also started experimenting with {\it attractive} headlines to catch readers' attention. These organizations are deploying A/B tests by exposing different headlines to different readers to identify the one capturing the readers' attention most
\footnote{\url{nytimes.com/2016/06/13/insider/which-headlines-attract-most-readers.html}}.
It is to be seen whether such strategies can bring some of the clickbait consumers to the traditional news media landscape. Additionally, to have more engagements with social media audience, traditional media organizations can adopt some strategies from clickbait media's playbook. They can use more pictures and hashtags in their tweets. They can also utilize the user mentions and replies to establish direct connections with their intended audience. Such outreach can help in getting more clicks, and also reaching the users who feel left-out in the traditional news discourse.

Finally, in certain scenarios, clickbait prevention may be required. For example, there has been a recent development central to any discussion around news stories in social media -- `Fake News'~\cite{vargo2017agenda}. Debates are going on regarding the spread of many false and misleading articles in social media, and its role in shaping public opinion during election periods~\cite{allcott2017social}. To attract readers, such fake news articles often deploy catchy sensational headlines. Similarly, articles propagating extreme opinion also deploy clickbait techniques to get more views. In such situations, going beyond detecting clickbaits, we may need to distinguish between such `bad' vs potentially `good' clickbaits, and apply prevention techniques (e.g., methods proposed in~\cite{chakraborty2016stop}) to stop widening their reach.

The solutions proposed above are only a few possible options to tackle the lowering of news value with the prevalence of clickbaits. Further research is needed to explore other alternatives. The study of clickbaits is still in its nascent stage, and there are a lot of open questions for future pursuits. For example, conducting in person interviews with the most active consumers of both clickbaits and non-clickbaits will help in understanding the consumer's perspectives on the usefulness of clickbaits as well as traditional news.

Additionally, there is a need to compare more fine-grained user characteristics in future studies. For example, 
different business organizations maintain Twitter accounts to communicate with their consumers~\cite{saffer2013effects}. 
Such organizational accounts form a substantial group on Twitter. It might be interesting to compare the presence of organizational accounts among clickbait and non-clickbait consumers. The non-clickbait consumers may involve more organizational 
accounts, while the clickbait consumers may include more individual users. 
Similar to organizational accounts, in a recent study, Varol {\it et al.}~\cite{varol2017online} 
claimed that as much as $15\%$ of Twitter accounts may be bots rather than real people. It is also worth investigating, whether certain media outlets have many bots among their consumers, which in turn may help in boosting their popularity in the eyes of the real consumers.

In conclusion, before making any decision on the fate of clickbaits in the future, it is of paramount importance to get the holistic picture and assess its social implications. In this work, we have made the first attempt towards that direction, and we hope that it will trigger discussions  similar to what happened around the tabloidization in the offline media. 

\vspace{-2mm}
\bibliographystyle{ACM-Reference-Format}
\bibliography{Main} 

\end{document}